\documentclass[global,twocolumn]{svjour}
\usepackage{graphicx}

\journalname{}

\begin{document}
\title{A scalable quantum computer with an ultranarrow optical transition of ultracold neutral atoms in an optical lattice}

\author{K. Shibata\inst{1}, S. Kato\inst{1}, A. Yamaguchi\inst{1}, S. Uetake\inst{1,2} \and Y. Takahashi\inst{1,2}
}

\institute{Department of Physics, Graduate School of Science, Kyoto University, Kyoto 606-8502, Japan 
\and
CREST, Japan Science and Technology Agency, 4-1-8 Honcho Kawaguchi, Saitama 332-0012, Japan}
\date{Received: date / Revised version: date}
%
\maketitle
\begin{abstract}
We propose a new quantum-computing scheme using ultracold neutral ytterbium atoms in an optical lattice. 
The nuclear Zeeman sublevels define a qubit. 
This choice avoids the natural phase evolution due to the magnetic dipole interaction between qubits. 
The Zeeman sublevels with large magnetic moments in the long-lived metastable state are also exploited to address  individual atoms and to construct a controlled-multiqubit gate.
Estimated parameters required for this scheme show that this proposal is scalable and experimentally feasible.
\end{abstract}

\section{Introduction}
In recent years a variety of quantum computing schemes have been proposed and 
many experiments have been performed\cite{Roadmap}. 
According to DiVincenzo\cite{DiVincenzo}, a quantum computer must fulfill following 5 criteria: 
(1) a scalable physical system with well-defined qubits; 
(2) the ability to initialize the states of the qubits to a simple state; 
(3) long decoherence time compared to gate operation time; 
(4) a universal set of quantum gates; 
(5) a qubit-specific measurement capability. 
So far, a nuclear magnetic resonance (NMR) system with less than 10 qubits was used to demonstrate a simple quantum algorithm\cite{NMR}.  
Recently an ion trap quantum computer with 8 entangled qubits was realized\cite{Haffner}. 
A quantum computer with neutral atoms in an optical lattice, so called an optical lattice quantum computer, is very attractive in view of the scalability.
More than $10^4$ atoms can be loaded into an optical lattice, 
and each atom confined in the individual site of the optical lattice\cite{Greiner} is regarded as a qubit. A special class of multi-partite entangled state, so called a cluster state, was successfully created via controlled collision in an optical lattice\cite{Mandel,Schrader}. Quite recently, controlled exchange interaction between pairs of atoms in an optical lattice has been demonstrated\cite{Anderlini,Trotzky}, which leads to a quantum SWAP gate.

One of the major difficulties associated with an optical lattice quantum computer scheme is, however, the individual addressing of a single qubit. 
The distance between adjacent atoms in an optical lattice is usually
about several 100 nm corresponding to a half of the wavelength of the laser for an optical lattice, which is usually very difficult to well resolve spatially by optical means. Note that imaging single atoms in a 3D optical lattice with the lattice spacing of several micrometer was successfully demonstrated\cite{Weiss}. 
In such a long-distance optical lattice, however, one needs to introduce an appropriate strength of interaction between atoms in adjacent lattice sites by some means, which is not yet demonstrated so far\cite{BlockadeEffect}.  
A different type of promising approach is to utilize a spectral addressing\cite{Cory,Demille,Derevianko}. 
The gradient of an electric or magnetic field over optical lattice sites can introduce a site-dependent resonance energy, and thus allows one to address an individual qubit spectroscopically, if the atoms or molecules have an appropriate strength of a magnetic or electric dipole moment. Such a technique has been successfully applied to the Rb Bose-Einstein condensate (BEC) to reveal a shell structure of a Mott-insulator state\cite{MRI-Bloch}. However, the resolution is not high enough to resolve individual sites so far.

\begin{figure}
\begin{center}
\includegraphics[width=9cm]{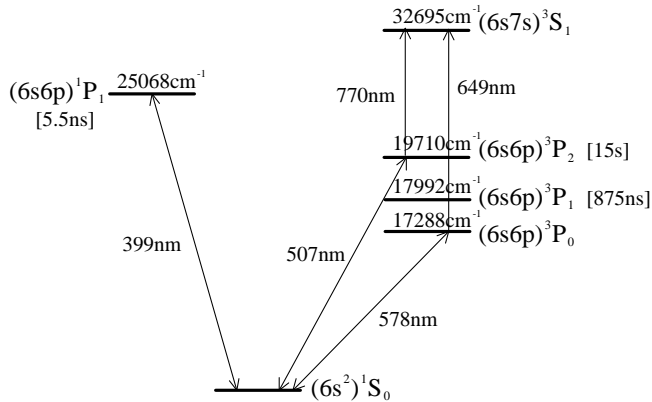}
\end{center}
\label{fig:Ybenergylevel}
\caption{Energy levels of ytterbium. Levels associated with quantum computation are shown.}
\end{figure}

In this paper, we propose a new scheme of an optical lattice quantum computer using a fermionic isotope of ytterbium $^{171}$Yb which has nuclear spin 1/2\cite{QITetc}. 
Yb is an alkali-earth like atom and has the energy levels shown in Fig. \ref{fig:Ybenergylevel}.
By exploiting the unique properties of various optical transitions and the energy levels of the $^{171}$Yb atom, we can construct a quantum computer with many advantageous features. 
In particular, our proposal is different from other proposals, especially Ref. \cite{Derevianko}, in that we consider two different kinds of qubits, one composed of the nuclear spin 1/2 for memory and the other of the electron spin for a gate operation. 
For connecting these two qubits as well as for the spectral addressing we use the ultranarrow optical transition and thus we realize an addressing of individual sites in an optical lattice as well as a hardware switching of inter-atomic interaction. 
The hardware switching of the interaction is especially important for a scalable quantum computer, otherwise the interaction is "always on", 
and the associated phase evolution of the system should be canceled by applying many short pulses, the number of which increases proportionally to the square of the number of qubits\cite{Leung}. In addition, our scheme naturally implements a single qubit measurement with an already well-demonstrated technique\cite{single atom detection}.
The details of these schemes are explained in Sec. \ref{sec: proposed scheme}.

\section{Proposed scheme}\label{sec: proposed scheme}

\subsection{ energy levels and optical transitions of $^{171}$Yb }
First we explain the unique properties of the energy levels and the optical transitions of the $^{171}$Yb atom, 
as shown in Fig. \ref{fig:Ybenergylevel}, and define two different kinds of qubits.
In the ground state $(6s^2)^1$S$_0$ there is no electron spin and only the nuclear spin 1/2 exists. The magnetic moment associated with the nuclear spin is 0.49367$\mu_N$ which is about  three-orders-of-magnitude smaller than that with an electron spin. The dipole-dipole interaction between the nuclear spin in an optical lattice of 266 nm lattice constant, for example, is 50 nHz, which is negligibly small for a typical experimental time of several seconds. 

We take this nuclear spin $1/2$ $\vert m_I=+1/2 \rangle$ and $\vert m_I=-1/2 \rangle$ as a qubit for memory.
To initialize the qubit, the optical pumping technique can be used. In addition, the atoms are loaded into the optical lattice with one atom in one site as a band insulator of polarized fermions\cite{koel}.
\begin{figure}
 \begin{center}
 \includegraphics[width=8cm]{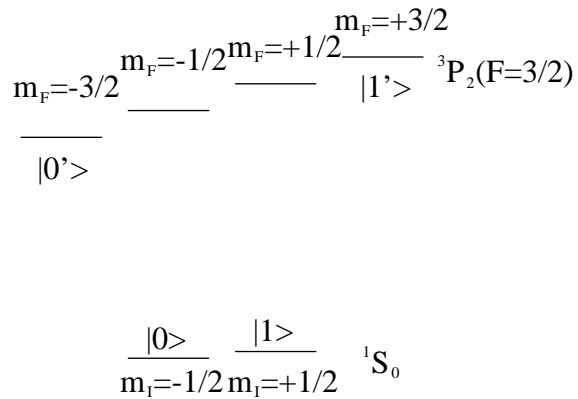}
 \end{center}
\caption{The Zeeman sublevels of a $^{171}$Yb in the $^1$S$_0$ and $^3$P$_2(F=3/2)$ states. Choice of computational basis and auxilliary states.}
\label{fig:Fermion scheme}
\end{figure}
This choice of the qubit avoids natural phase evolution due to the interaction between qubits, while 
it does not allow us individual addressing and controlled-multiqubit gate operations. 

In the long-lived metastable $(6s6p)^3$P$_2$ state with a lifetime of 15 secs, on the other hand, there is a large magnetic moment of $3\mu_B$ associated with the electron spin and orbital angular momentum. 
This large magnetic moment is advantageous to construct a controlled-multiqubit gate\cite{Derevianko}. 
In fact, if two atoms in the $^3$P$_2$ state are arranged along the quantization axis in an optical lattice of 266 nm lattice constant,
the dipole-dipole interaction between the two adjacent atoms is 10 Hz , which is large enough to perform many gate operations for a typical experimental time of several seconds. We consider the Zeeman sublevels $\vert m_F=+3/2 \rangle$ and $\vert m_F=-3/2 \rangle$ of the hyperfine structure F=3/2 in the $^3$P$_2$ state as an auxiliary qubit for a gate operation.

The optical transition between the ground state $^1$S$_0$ and the metastable state $^3$P$_2$ is only weakly allowed by hyperfine-interaction-induced electric dipole (E1) transition\cite{Porsev} and thus the linewidth is as narrow as 10 mHz. Owing to this ultranarrow linewidth and the large magnetic moment of the $^3$P$_2$ state, 
we can obtain a spatial resolution of 266 nm for a modest strength of a magnetic field of 10 G/cm and a spectral resolution of 1 kHz. 
In addition, by coherently transferring the atomic states between the $^1$S$_0$ and $^3$P$_2$ states using this transition, we can realize a switching of the magnetic dipole-dipole interaction, which is the key issue to construct quantum gates in a scalable manner. 

The transition between $^1$S$_0$ and the $(6s6p)^1$P$_1$ state ($^1$P$_1$) is, on the other hand, a strongly-allowed E1 transition,  and the $^1$P$_1$ state has a very short lifetime of 5.5ns\cite{Porsev2}. Thus, this transition is quite useful for the measurement of a single qubit with an already well-demonstrated fluorescence detection technique from a magneto-optical trap \cite{single atom detection}.

\subsection{individual addressing}
The gradients of a magnetic field are used for the individual addressing\cite{Derevianko}. 
We choose the z axis as the quantization axis, and apply a strong magnetic field $B_0$ along the z direction.
First we apply a field gradient along the z direction $\frac{\partial B_z}{\partial z}$ to select an only one particular layer of a x-y plane in a 3D optical lattice. Only the atoms in the selected layer is transferred to the $^3$P$_2$ state by the laser resonant to the $^1$S$_0 \leftrightarrow ^3$P$_2$ transition, and the other atoms remaining in the $^1$S$_0$ state are blurred away from the optical lattice by the strong radiation pressure associated with the transition $^1$S$_0 \leftrightarrow ^1$P$_1$. Then we return the atoms in the  $^3$P$_2$ state to the $^1$S$_0$ by the laser resonant to the $^1$S$_0 \leftrightarrow ^3$P$_2$ transition again. This is how we can prepare the atomic ensemble of an only one layer of the x-y plane.

\begin{figure}
 \begin{center}
  \includegraphics[width=3.5cm]{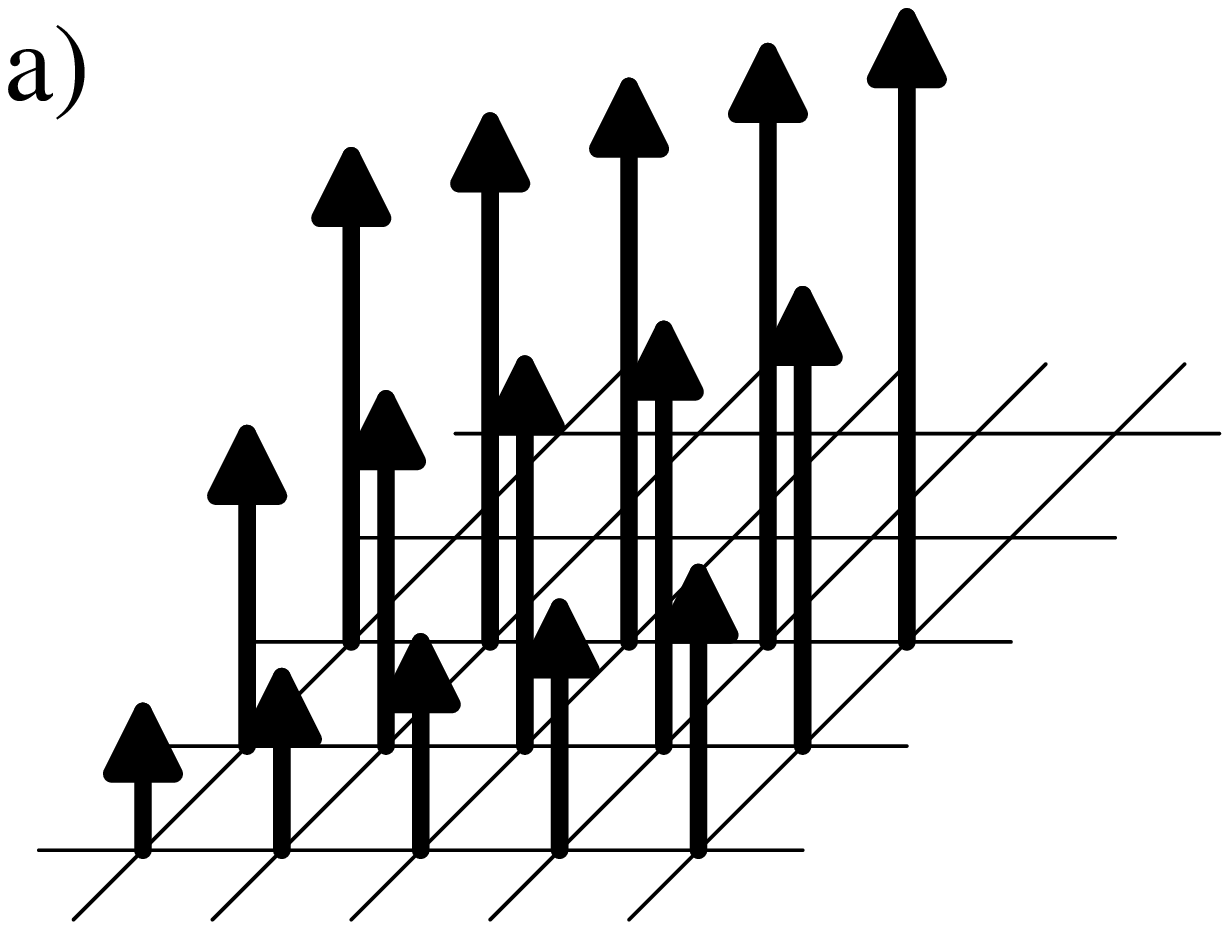}
 \hspace{1mm}
  \mbox{\raisebox{4mm}{\includegraphics[width=4.6cm]{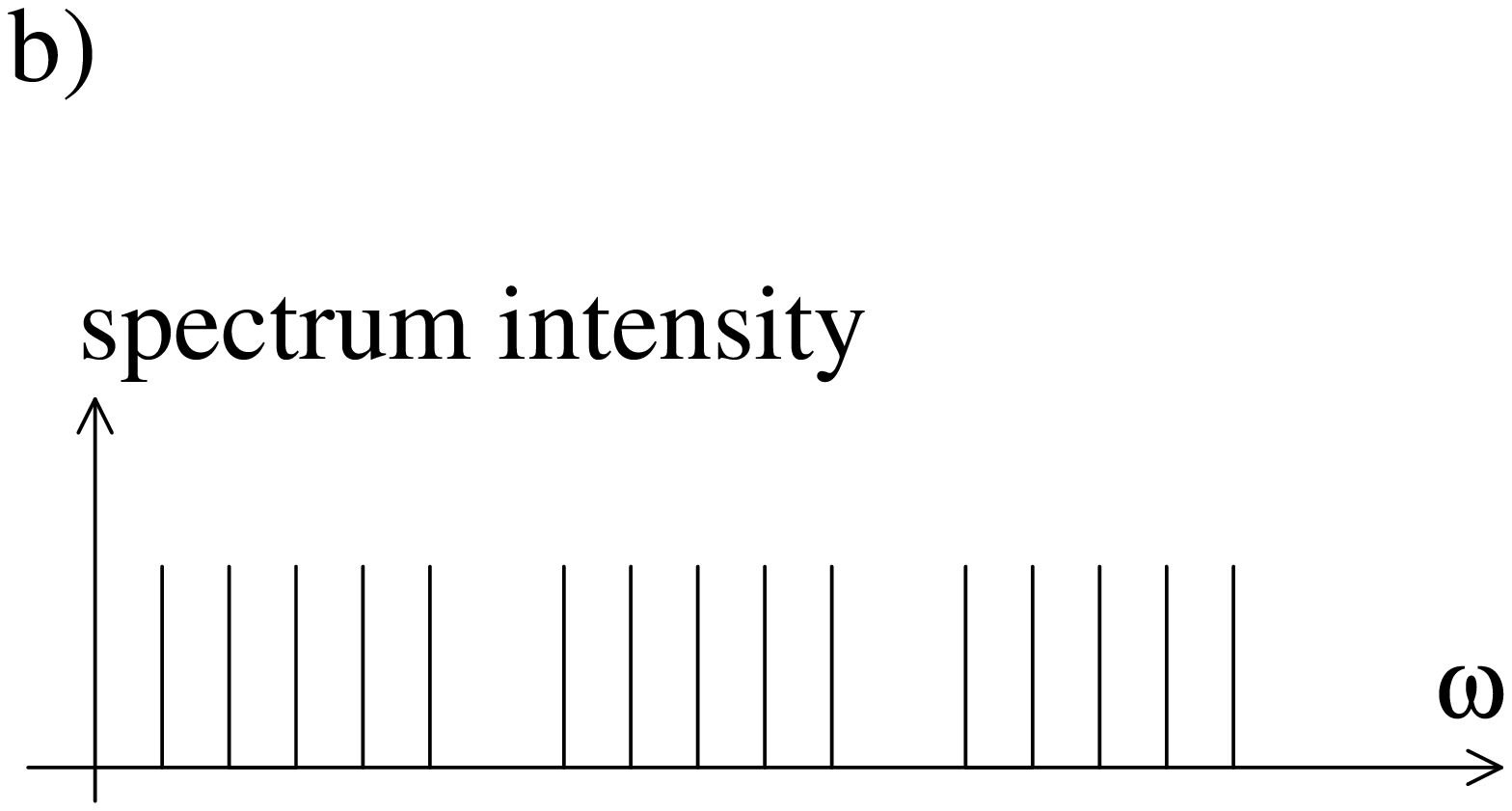}}}
  \end{center}
  \caption{Principle of individual addressing. Arrows on each site in the left figure represent magnetic fields on the site. 
  The right figure shows an expected resonant spectrum.}
  \label{image of individual adressing}
\end{figure}

Next we apply the field gradients along the x and y directions
$\frac{\partial B_z}{\partial x}$ and $\frac{\partial B_z}{\partial y}$, which results in position-dependent Zeeman shifts and  enables one to resolve two neighboring atoms in the  x-y plane.
Note that $\frac{\partial B_z}{\partial x}$ and  $\frac{\partial B_z}{\partial y}$ must satisfy 
\begin{equation}\label{gradients relation}
n_x \times \frac{\partial B_z}{\partial x} \leq \frac{\partial B_z}{\partial y}, 
\end{equation}
so that the magnetic field on each site differs from each other. 
(See Fig. \ref{image of individual adressing}.) 
Here $n_x$ is the number of atoms along the x-axis. 

\subsection{single qubit gate}
\begin{figure}
\begin{center}
 \mbox{\raisebox{2.5mm}{\includegraphics[height=3.8cm]{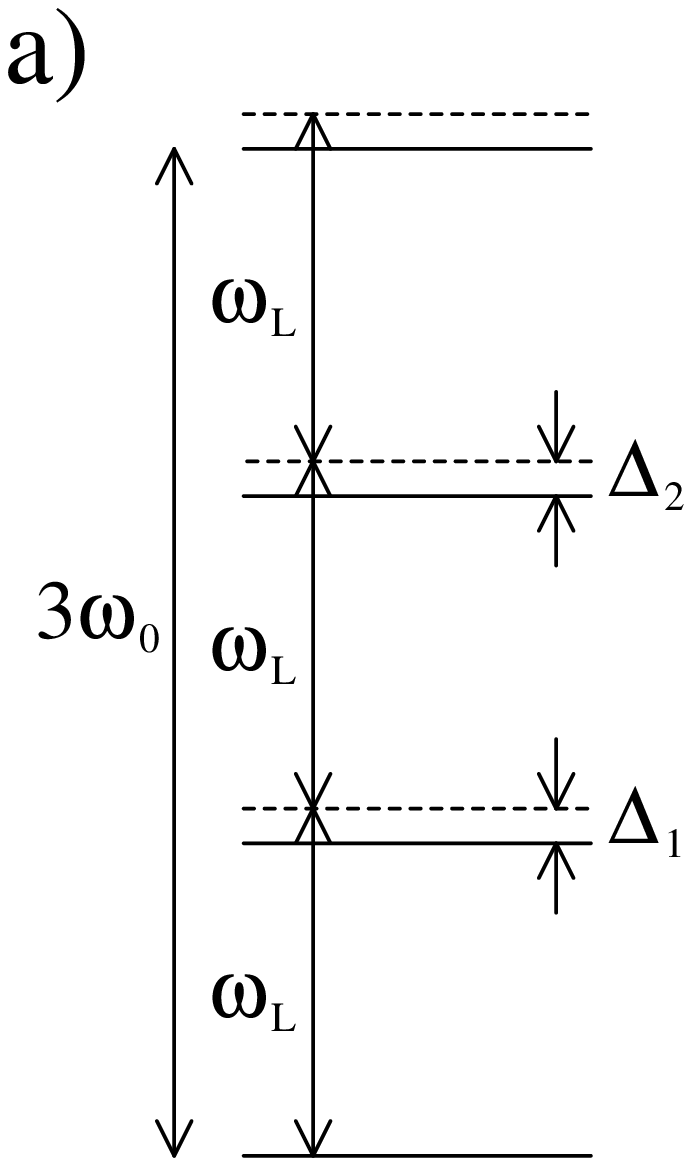}}}
 \includegraphics[height=4cm]{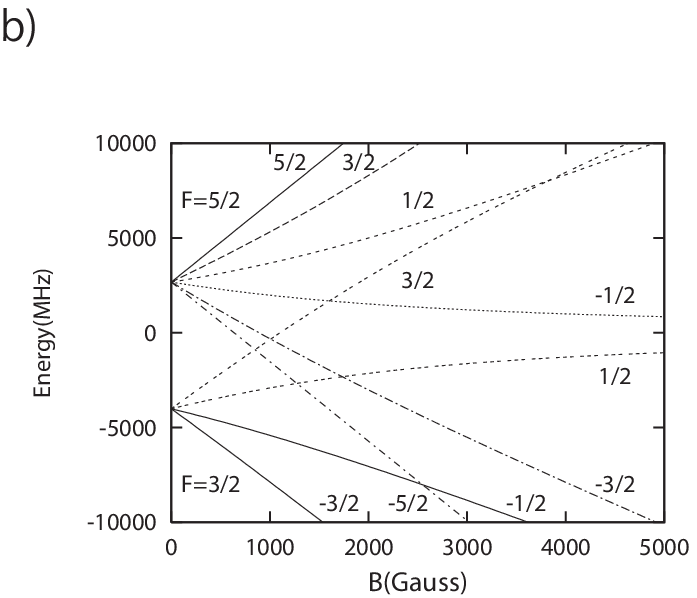}
\end{center}
\caption{(left)A four level system. Dashed lines are virtual states and solid lines are real states.
(right)Zeeman levels of the $^3$P$_2$ state of $^{171}$Yb.}
\label{fig:3 photon transition}
\end{figure}

A global single qubit unitary operation can be performed for the qubits in the $^1$S$_0$ state by a conventional NMR technique. A specific single qubit unitary operation can be performed after the atomic state in the qubit is coherently transferred to the $^3$P$_2(F=3/2)$ state by two laser fields resonant to the $^1$S$_0(m_I=+1/2) \leftrightarrow ^3$P$_2(F=3/2, m_F=+3/2)$ and $^1$S$_0(m_I=-1/2) \leftrightarrow ^3$P$_2(F=3/2, m_F=-3/2)$transitions in an appropriate field gradient. 
In the $^3$P$_2(F=3/2)$ state, 3 photon Rabi oscillation in the four level system can be exploited as shown in Fig. \ref{fig:3 photon transition}. 

\begin{figure}[b]
 \begin{center}
  \includegraphics[width=6cm]{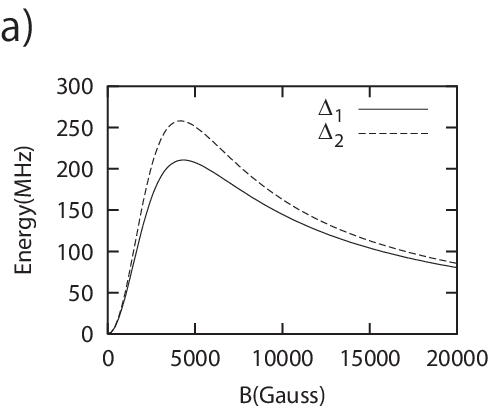}
  \includegraphics[width=6cm]{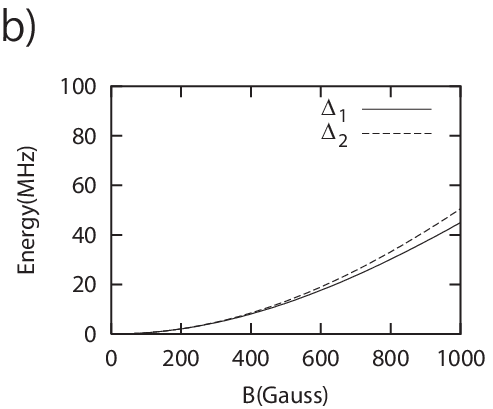}
 \end{center}
 \caption{The top graph shows calculated values of $\Delta_1$ and $\Delta_2$ for the magnetic field of 0 to 20000 Gauss. 
The bottom graph shows calculated values of $\Delta_1$ and $\Delta_2$ for the low magnetic field (0-1000 Gauss). 
Solid lines and dashed lines in these graphs represent $\Delta_1$ and $\Delta_2$ respectively. }
 \label{fig:Delta1&Delta2}
 \end{figure}

When the detunings $\Delta_1= \omega_{ab}-\omega_0$ and $\Delta_2= \omega_{cd}- \omega_0$ are much larger than the Rabi frequency $\Omega$ of single photon transition between the states a and b, b and c, and c and d, the atom oscillates between the state a and state d
with the frequency $\Omega_{eff}^{(3)}=\frac{\Omega^3}{4\Delta_1\Delta_2}$ \cite{Shore}. 
Here $\omega_{ab}$ and $\omega_{cd}$ represent $(E_{b}-E_{a})/\hbar $ and  $ (E_{d}-E_{c}) /\hbar $, repectively, and  a, b, c, and d represent the $m_F= -3/2, -1/2, +1/2, $ and $+3/2$, respectively.
To evaluate $\Omega_{eff}^{(3)}= \Omega^3/4\Delta_1\Delta_2$, 
we first calculate $\Delta_1$ and $\Delta_2$. 
Calculated values of $\Delta_1$ and $\Delta_2$ are shown in Fig. \ref{fig:Delta1&Delta2}.

Unless $\Delta_1$ and $\Delta_2$ are much larger than $\Omega$, the transition to intermediate states occurs. 
Therefore, we need large $\Delta_1$ and $\Delta_2$. 
For example, at B=650Gauss $\Delta_1$ and $\Delta_2$ are about 20MHz and 
a 1ms operation with error probability p=0.01 is realized when $\Omega=2\pi \times 985kHz$. 
Note that the non-zero value of $\Delta_{1(2)}$ comes from the nonlinearity of the Zeeman shift due to the coupling between nuclear spin and electronic angular momentum. (See Fig. \ref{fig:3 photon transition}.)

After the 3 photon Rabi oscillation, the atom of the auxiliary qubit state is brought back again to the ground state $^1$S$_0$ similarly by two laser fields resonant to the $^1$S$_0(m_I=+1/2) \leftrightarrow ^3$P$_2(F=3/2, m_F=+3/2)$ and $^1$S$_0(m_I=-1/2) \leftrightarrow ^3$P$_2(F=3/2, m_F=-3/2)$ transitions with preserving the spin coherence. The other advantageous aspect of this scheme is that the decoherence of the qubit is not limited by a finite lifetime of the $^3$P$_2$ state differently from the scheme in Ref.\cite{Derevianko}.

\subsection{CNOT gate}
\begin{figure}[b]
 \begin{center}
  \includegraphics[width=8cm]{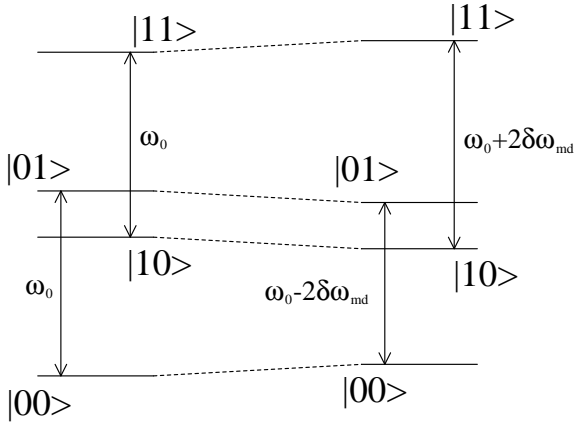}
 \end{center}
 \caption{(left) Energy levels excluding magnetic dipole interaction. 
 (right) Energy levels including magnetic dipole interaction.}
 \label{fig:CNOT image}
 \end{figure}

A specific two-qubit CNOT operation can be performed similarly after the two-adjacent qubits are selectively transferred to the $^3$P$_2(F=3/2)$ state by the laser resonant to the $^1$S$_0 \leftrightarrow ^3$P$_2(F=3/2)$ transition in the field gradient. 
Then a CNOT gate operation is performed between the qubits in the auxiliary states $\vert m_F=+3/2 \rangle$ and $\vert m_F=-3/2 \rangle$. 
The resonant frequencies of the transitions $\vert 00 \rangle \leftrightarrow \vert 01 \rangle$ and $\vert 10 \rangle \leftrightarrow \vert 11 \rangle$ differ by 40 Hz in an optical lattice of 266 nm lattice constant due to the magnetic-dipole interaction (See Fig. \ref{fig:CNOT image}.). The application of a $\pi$-pulse resonant to $\vert 10 \rangle \leftrightarrow \vert 11 \rangle$ transition results in the CNOT gate. Here 0 and 1 represent the state with $m_F=-3/2$ and $m_F=+3/2$, respectively.  
After the CNOT operation, as in the case of the single qubit operation, the atoms of the auxiliary qubit states are brought back again to the ground state $^1$S$_0$ by two laser fields resonant to the $^1$S$_0(m_I=+1/2) \leftrightarrow ^3$P$_2(F=3/2, m_F=+3/2)$ and $^1$S$_0(m_I=-1/2) \leftrightarrow ^3$P$_2(F=3/2, m_F=-3/2)$ transitions. In this way we turn on and off the magnetic dipole-dipole interaction.

\subsection{measurement}
A single qubit measurement is usually difficult for an optical lattice based quantum computing scheme. 
In Ref. \cite{Derevianko} a state-selective ionization is the candidate for measurement of each qubit. 
Our scheme naturally implements a single qubit measurement with an already well-demonstrated technique of the fluorescence detection from a magneto-optical trap (MOT) \cite{single atom detection} in combination with the individual addressing technique already mentioned above.
First, we transfer all the qubit in the ground state $^1$S$_0$ to the  $^3$P$_2$ state by laser fields in the absence of the field gradient. Note that at this measurement stage we only concern the probability of the qubit state and do not care about any phase evolution due to the magnetic dipole-dipole interaction in the $^3$P$_2$ state. 
Second, the atomic state $^3$P$_2(F=3/2, m_F=+3/2) $ or $^3$P$_2(F=3/2, m_F= -3/2)$ of the particular qubit of interest is brought back again to the ground state $^1$S$_0$ similarly by the laser field resonant to the $^1$S$_0(m_I=+1/2) \leftrightarrow ^3$P$_2$(F=3/2, m$_F$=+3/2) or 
 $^1$S$_0$(m$_I$=-1/2) $\leftrightarrow$ $^3$P$_2$ (F=3/2, m$_F$=-3/2) transition.
Then, finally we apply the laser fields resonant to the transition $^1$S$_0$ $\leftrightarrow$ $^1$P$_1$(F=3/2) for the MOT and detect the fluorescence from the MOT. Due to the short lifetime of the $^1$P$_1$ state, we expect the fluorescence count rate high enough for the single atom detection in a usual light-collecting setup. If we detect the fluorescence or do not detect the fluorescence, the qubit state is then collapsed to the corresponding state. 
We repeat this sequence to detect all of the qubits.
We note that this measurement scheme fails for a long fluorescence detection time due to the existence of the small branching from the $^1$P$_1$ state to the $(5d6s)^3$D$_1$ and $(5d6s)^3$D$_2$ states. This effect, however, is negligible for our typical measurement time of several milli seconds.

\section{Experimental Feasibility}\label{sec: feasibility}
Here we discuss the experimental feasibility of our proposal. 
$^{171}$Yb atoms are successfully cooled to ultracold temperature of about 100nK with sympathetic cooling method, and also loaded into the optical lattice of 266 nm lattice constant. Moreover, quite recently the ultranarrow optical transition $^1$S$_0 \leftrightarrow ^3$P$_2$ was successfully observed for $^{171}$Yb and $^{174}$Yb atoms\cite{private}. The necessary laser intensity for a 100$\mu$s $\pi$-pulse for the $^1$S$_0 \leftrightarrow ^3$P$_2$ transitions is $4.82\times 10^4 W/m^2$, which is easily obtained experimenally.
Therefore we do not expect great difficulty in the preparation of the cold atoms and the optical excitations. 
To suppress the tunneling between adjacent optical lattice sites in a typical experimental time of 5 seconds, for example, the potential depth should be about 50 times larger than the recoil energy, which corresponds to about 10 $\mu$K depth for the lattice of 266 nm lattice constant. The photon scattering rate at this lattice depth is about 1/5 Hz, thus this would not be a fatal problem. 

The implementation of stable magnetic field gradients in the x-y plane would be the most important technical issue in our proposal. The necessary strength of the magnetic field gradients are evaluated by the condition that any two resonant frequencies should be resolved. 
To obtain the frequency difference of 1 kHz for $n_x=10$ and $n_y=10$ lattice, 
$\frac{\partial B_z}{\partial x}$ = 10 G/cm and $\frac{\partial B_z}{\partial y}$= 100 G/cm
as well as $\frac{\partial B_z}{\partial z}$ = 100 G/cm are sufficient. 
A bias magnetic field $B_0$ is applied to define the quantization axis besides these gradients. 
$B_0=100G$ is sufficient in a system with less than 1000 qubits. 
Such field gradients and $B_0$ can be prepared by using some coils such as a Helmholtz coil, an anti- Helmholtz coil, and saddle coils. These coils should be surrounded by a magnetic shield to avoid stray magnetic fields. In addition, especially important is the spatial stability of the fields relative to the optical lattices. The monolithic design of the field generating coils and the optical lattices would be effective for this purpose.

It is also true that this coil configuration has difficulty for compatibility with a standard setup of cold atom experiments. To spatially transfer cold atoms produced in a standard setup of cold atom experiment into a small glass cell which are compatible with the coil configuration, the well demonstrated technique of a moving optical tweezer\cite{tweezer} is expoited. 


\section{Conclusion}
We proposed a new scheme for quantum computation using $^{171}$Yb atoms in an optical lattice. 
A quantum computer based on our scheme fulfills DiVincenzo's criteria; 
(1) scalability: more than 1000 atoms can be loaded into an optical lattice. 
(2) initialization: the initialization of the qubits is done by cooling and optical pumping. 
(3) long decoherence time compared to operation time: the ground state $^1$S$_0$ would offer long decoherece time of several seconds and many operations are performed in this experimental time with realistic parameters. 
(4) a universal set of quantum gates: the individual addressing is achieved with a spectral addressing with the ultranarrow intercombination transition $^1$S$_0 \leftrightarrow ^3$P$_2$ under a magnetic field gradient.
The single qubit gates are performed by 3 photon Rabi oscillations in the $^3$P$_2$ state, and the
magnetic dipole-dipole interaction also in the $^3$P$_2$ state is exploited in the CNOT gate operations. 
The interaction switching by controlling internal states of atoms enables us to construct the CNOT gates in a large system. 
(5) measurement: the fluorescence from a MOT is exploited to measure individual qubits after the individual addressing. 

Although we have specifically considered a gate-based quantum computation, the proposed scheme would be also applicable to measurement-based one-way quantum computation using cluster states\cite{cluster state}.


\begin{acknowledgement}
This work was partially supported by Grant-in-Aid for Scientific Research of JSPS (Grant No. 18204035) and the Global COE Program gThe Next Generation of Physics, Spun from Universality and Emergenceh from the Ministry of Education, Culture, Sports, Science and Technology (MEXT)
of Japan. 
S.K. acknowledges support from JSPS.
\end{acknowledgement}

\end{document}